\documentclass[runningheads]{llncs}
\usepackage[T1]{fontenc}
\usepackage{graphicx}
\usepackage{booktabs}
\usepackage[misc]{ifsym}
\newcommand{\corr}{(\Letter)}
\usepackage{multirow}
\usepackage{amssymb}
\usepackage{amsmath}
\usepackage{bbm}
\usepackage{slashbox}
\usepackage{enumitem}
\usepackage{svg}
\usepackage{xcolor}
\usepackage[hyphens]{url}

\begin{document}

\title{Integrating Optimal Transport and Structural Inference Models for GRN Inference from Single-cell Data}

\titlerunning{Integrating Optimal Transport and Structural Inference}

\author{
Tsz Pan Tong\inst{1,2}\orcidID{0000-0001-8111-5886}~\corr \and
Aoran Wang\inst{1}\orcidID{0000-0001-7809-0622} \and
George Panagopoulos\inst{1}\orcidID{0000-0001-7731-9448} \and
Jun Pang\inst{1,2}\orcidID{0000-0002-4521-4112}
}

\authorrunning{T.~P. Tong et al.}

\institute{Department of Computer Science, University of Luxembourg, Luxembourg \and Institute for Advanced Studies, University of Luxembourg, Luxembourg \\
\email{\{tszpan.tong,aoran.wang,georgios.panagopoulos,jun.pang\}@uni.lu}}

\maketitle              

\begin{abstract}
We introduce a novel gene regulatory network (GRN) inference method that integrates optimal transport (OT) with a deep-learning structural inference model.
Advances in next-generation sequencing enable detailed yet destructive gene expression assays at the single-cell level, resulting in the loss of cell evolutionary trajectories.
Due to technological and cost constraints, single-cell experiments often feature cells sampled at irregular and sparse time points with a small sample size.
Although trajectory-based structural inference models can accurately reveal the underlying interaction graph from observed data, their efficacy depends on the inputs of thousands of regularly sampled trajectories.
The irregularly-sampled nature of single-cell data precludes the direct use of these powerful models for reconstructing GRNs.
Optimal transport, a classical mathematical framework that minimize transportation costs between distributions, has shown promise in multi-omics data integration and cell fate prediction.
Utilizing OT, our method constructs mappings between consecutively sampled cells to form cell-level trajectories, which are given as input to a structural inference model that recovers the GRN from single-cell data.
Through case studies in two synthetic datasets, we demonstrate the feasibility of our proposed method and its promising performance over eight state-of-the-art GRN inference methods.

\end{abstract}

\section{Introduction}
\label{sec:intro}
Genes store the information necessary for the development and functioning of living organisms.
Gene regulation is the process of controlling the expression of genes, which are essential for cell growth, development, and maintenance.
Understanding gene regulations is crucial for elucidating the mechanisms underlying biological processes and diseases.
Biologists conduct single-cell RNA sequencing (scRNA-seq) experiments, in which specific types of cells are grown and left to evolve under a carefully controlled environment to record a targeted biological process.
During the experiment, portions of cells are killed and assayed to obtain the readings of gene expression levels for each cell at designed time steps.
Gene expression data is used to infer the gene regulatory network (GRN), a mathematical model representing gene regulations.
Over the past decade, the biology community has accumulated a wealth of scRNA-seq data, which motivates the development of various computational methods for inferring GRNs from single-cell datasets.

Currently, inferring GRN from single-cell data faces several challenges.
First, the assay of scRNA-seq data is destructive, meaning that the gene expression levels of each cell can only be measured at a single time point, resulting in the loss of cell evolutionary trajectories.
Assays at each time point only provide a snapshot of gene expression distribution.
Second, single-cell experiments are expensive. 
Cells are often sampled in small batches at irregular and sparse time points, and critical changes in the gene regulation dynamics are often missed in observations.
Third, a complete GRN includes a large number of genes, which makes the inference problem high-dimensional.
Limiting the size of the gene set will inevitably introduce confounding variables that may lead to false discovery.
Finally, experimentally validated GRNs are scarce, which makes it even more challenging to evaluate the performance of GRN inference methods.

In addition, structural inference is a fast-growing field that aims to infer the underlying interactions between agents in a dynamical system from observed data, especially in the era of deep learning.
Compared with classical GRN inference methods, deep learning-based models have customizable architectures that allow the adoption of various forms of input data and model assumptions.
This grants greater flexibility in model design and empowers the model to capture the underlying structure of the dynamical system more accurately.
Kipf et al.~\cite{kipf2018nri} proposed the neural relational inference (NRI) model that uses graph neural networks (GNNs) to infer the interaction network of a dynamical system.
Its data-driven approach, message-passing mechanism, and GNN architecture match some of the needs of GRN inference.
A recent benchmarking study~\cite{StructInfer} shows NRI and its variants achieve high inference accuracy in various simulated dynamical systems.
In addition, the NRI model can output a directed graph with edge probabilities, which avoids the need to set a threshold to determine the existence of an edge in classical GRN inference methods.
However, the NRI's accurate performance relies on thousands of regularly and frequently sampled trajectories, which excludes its direct use in GRN inference.
Thus, reconstructing cellular trajectories from snapshots of gene expression distributions is critical in applying NRI to GRN inference.

Optimal transport (OT) is a classical mathematical problem that aims to find a mapping between elements in two distributions that minimizes transportation cost, which is an ideal tool for reconstructing cellular trajectories.
OT has been successfully applied in various fields, such as image processing~\cite{perrot2016mapping,titouan2020coot}, economics~\cite{fajgelbaum2020optimal}, and biology~\cite{yang2020predicting}.
In single-cell data analysis, OT has been used to integrate multi-omics data~\cite{cao2022unified,huizing2022optimal}, model cell dynamics~\cite{tong2020trajectorynet,bunne2022jkonet} and predict cell fate~\cite{schiebinger2019wot}.
Motivated by the successful applications of OT, we propose a novel method for inferring GRN from single-cell data that integrates OT with the NRI model.
We reconstruct cell evolutionary trajectories using OT, which minimizes the transportation cost between gene expression distributions at consecutive time points.
The reconstructed trajectories are then used as inputs to the NRI model to infer the underlying GRN.
In this way, our approach combines the strengths of OT and structural inference models and brings a new deep-learning tool into single-cell analysis.

\medskip
\noindent{\bf Contributions.}
Our contributions in this paper can be summarized as follows:
\begin{itemize}[wide, leftmargin=*]
    \item In this work, we propose a novel approach integrating OT with structural inference models for GRN inference from single-cell data.
    \item We achieved 73.96\% and 81.39\% AUROC scores in the mCAD and VSC datasets, ranked top and third among popular GRN inference methods.
    \item Our approach demonstrates the feasibility of using a broad range of structural inference models in GRN inference, which can be further customized to fit the features and dynamics of single-cell data.
    \item We analyze the difficulties faced in cell-level trajectory reconstruction and structural inference. 
\end{itemize}

\section{Related Work}
\label{sec:related}
\subsection{Gene Regulatory Networks Inference}
\label{ssec:grn}
GRN inference is difficult due to data scarcity, technical noise~\cite{jiang2022statistics}, and the complexity of biological systems. 
Biologists explore statistical methods for GRN inference because of its stability and explainability.
Information theory-based methods use mutual information~\cite{margolin2006aracne,faith2007clr}, information decomposition~\cite{chan2017pidc} and causal inference~\cite{qiu2020scribe,deshpande2022singe} to search for and screen out regulation pairs.
Correlation-based methods~\cite{kim2015ppcor,specht2017leap} use different correlation quantification methods as features to model the regulation relationship.
Tree-based machine learning methods~\cite{huynh2010genie3,moerman2019grnboost2,ma2020grnnonlinearode} use different tree algorithms to model the dynamic of cell evolution directly.
Their learned importance scores can be used to infer the regulatory relationship.
Other methods include regression~\cite{haury2012tigress,papili2018sincerities}, ordinary differential equation (ODE) modelling~\cite{matsumoto2017scode,aubin2020grisli}, matrix/tensor factorization~\cite{duren2018couplednmf,osorio2020sctenifoldnet}, time series modelling~\cite{sanchez2018grnvbem} and in-silico knockout~\cite{kamimoto2023celloracle}.
However, only a few models can incorporate the temporal information in the single-cell data and give a directed graph as the output, and there is a large model performance discrepancy between the use of synthetic and real-world datasets~\cite{zhao2021comprehensive}.
Interested readers can refer to various GRN benchmarking papers~\cite{pratapa2020benchmarking,nguyen2021comprehensive,zhao2021comprehensive} for more details.

Recently, deep learning methods have also played a role in GRN inference.
Thanks to intensive research in image processing, convolutional neural networks (CNN) are often used in model design.
These CNN-based models~\cite{yuan2019cnnc,xu2022dyndeepdrim,reagor2023delay} treat pairwise joint distributions of gene expression as images and use convolutional layers to predict the existence of regulation.
Others include time series models~\cite{zhang2019ngnc,Monti2022dadnn}, variational autoencoder (VAE) model~\cite{shu2021deepsem} and explainable AI model~\cite{keyl2023lrp}.
Deep learning models have shown promising results in GRN inference.
In contrast to statistical models, deep learning models offer a distinct advantage. They do not necessitate strong model prior assumptions, and their flexibility allows researchers to design customized architectures that can effectively capture the unique features of the data and the underlying dynamics.
However, deep learning models lack interpretability, and their performance is highly dependent on the quality and quantity of the training data.

\subsection{Structural Inference}
\label{ssec:si}
Structural inference is a field aimed at inferring agent interaction's underlying topological structure in a dynamical system.
NRI~\cite{kipf2018nri} is a representative of the deep learning model in structural inference, which has inspired a series of works using the VAE architecture.
NRI has several variants that improve the accuracy and training efficiency of the model.
iSIDG~\cite{wang2022isidg} iteratively updates the adjacency matrix by using the learned adjacency matrix from the previous iteration, creating a tighter information bottleneck in the VAE model and thus achieving a better performance.
RCSI~\cite{wang2023rcsi} integrates iSIDG with reservoir computing, which helps predict gene expression in the next time step.
This design significantly increases the training efficiency and thus requires fewer trajectories and fewer time steps to achieve the same performance as the original iSIDG model.
These models have shown their capability in inferring the underlying structure of various dynamical systems from synthetic data~\cite{StructInfer}, including synthetic GRN and gene coexpression data.
However, these models' outstanding performance depends on the abundance (around 12,000) of frequently and even sampled trajectories, and it is infeasible to collect such scRNA-seq datasets for GRN inference.

This paper proposes a novel method that integrates the OT-based trajectory reconstruction method with structural inference models to bridge the data requirement gap between structural inference models and GRN inference with single-cell data.
\section{Preliminaries}
\label{sec:pre}
In this section, we introduce the notations and two main tools used in our method: Waddington-OT (WOT)~\cite{schiebinger2019wot} for trajectory reconstruction and NRI for structural inference.
\subsection{Notations and Problem Definition}
For a scRNA-seq dataset sampled at $k$ time points $t_0,\dots, t_{k-1}$, we denote the number of genes as $g$ and the number of cells at time $t_i$ as $c_i$ ($0\le i \le k-1$).
The gene expression matrix at time $t_i$ is denoted as $X^{t_i}\in\mathbb{R}^{g\times c_i}$, where $X^{t_i}_{r,p}$ is the genes expression levels of gene $r$ in cell $p$.
$X^{t_i}_{\cdot,p}$ is the gene expression vector of cell $p$ at time $t_i$, where $\cdot$ denotes all value across a dimension.
The cell evolutionary trajectory $V_p=\{V^{t_i}_p\}_{i=0}^{k-1}$ in $\mathbb{R}^g$ is a sequence of gene expression values of cell $p$ following the chronological order. 

Structural inference treats GRN as a directed graph $\mathcal{G}=(\mathcal{V}, \mathcal{E})$, where $\mathcal{V}$ is the set of $g$ nodes (genes) and $\mathcal{E}\subset\mathcal{V}\times\mathcal{V}$ is the set of edges (regulatory relationship).
The adjacency matrix $A\in\mathbb{R}^{g\times g}$ is a matrix representation of the regulatory graph $\mathcal{G}$, where $A_{i,j}=1$ if gene $i$ regulates gene $j$, and $A_{i,j}=0$ otherwise.
Our research problem is to find the optimal adjacency matrix that best represents the underlying GRN from the observed scRNA-seq data.
%
%
\subsection{Waddington-OT (WOT)}
\label{ssec:trajectory_reconstruction}
Reconstructing cellular trajectories is beneficial for single-cell analysis on cell dynamics, lineage tracing, cell subtype identification and cell fate prediction.
In our case, we can utilize trajectory reconstruction tools to generate cell evolutionary trajectories as inputs to the structural inference model.
Many trajectory inference methods are proposed for lineage tracing and pseudotime estimations~\cite{saelens2019comparison} at the resolution of cell subtype level~\cite{farrell2018zebrafish} only.

WOT~\cite{schiebinger2019wot} is the pioneer of using OT to reconstruct cell-level evolutionary trajectories. 
WOT gives a probabilistic mapping between ancestor and descendant cells in consecutive time points. 
Inspired by WOT, most cell-level trajectory reconstruction methods are based on the optimal transport theory using Neural ODE models~\cite{tong2020trajectorynet,huguet2022mioflow,sha2024tigon}, which give a continuous trajectory at any point in the gene space. 
However, these Neural ODE trajectory reconstruction methods are hard to train and prone to over-smoothing trajectories.
The reconstructed trajectories never intersect because of the uniqueness of ODE solutions, which deviate from the cell's biochemical properties.

We use the WOT approach to reconstruct cell evolutionary trajectories.
For each pair of gene expression matrix $X^{t_i}, X^{t_{i+1}}$ at time $i,i+1$, we compute the pairwise cell distance matrix $M\in\mathbb{R}^{c_i\times c_{i+1}}$ as the cost matrix used in WOT.
We choose the Euclidean distance as the distance metric, where $M_{p,q}=\|X^{t_i}_{\cdot,p}-X^{t_{i+1}}_{\cdot,q}\|_2$ for cell $p$ at time $t_i$ and cell $q$ at time $t_{i+1}$. 
WOT solves the following unbalanced entropy-regularized OT optimization problem:

\begin{align}
    \gamma_{t_i\rightarrow t_{i+1}} = \operatorname*{arg\,min}_{\gamma\in\mathbb{R}^{c_i\times c_{i+1}}} \quad & \sum_{p,q} \gamma_{p,q}M_{p,q} + \epsilon \sum_{p,q} \gamma_{p,q}\log\gamma_{p,q} \\
    \textrm{subject to} \quad & \sum_p \gamma_{p,q} = \frac{1}{c_{i+1}} \quad \forall 1\leq q\leq c_{i+1} \nonumber \\
                        & \sum_q \gamma_{p,q} = \frac{1}{c_i}, \quad \forall 1\leq p\leq c_i \nonumber \\
                        & \gamma_{p,q}\geq 0 \quad \forall 1\leq p\leq c_i, 1\leq q\leq c_{i+1}, \nonumber 
\end{align}

where $\gamma_{t_i\rightarrow t_{i+1}}$ is the optimal transport plan between $X^{t_i}$ and $X^{t_{i+1}}$, and $\epsilon > 0$ is the entropic regularization term.

The optimization constraints are tackled by adding penalty terms in the objective function, thus the optimized $\gamma_{t_i\rightarrow t_{i+1}}$ may not row-wisely sum up to $\frac{1}{c_{i+1}}$.
WOT treats $\sum_q {\gamma_{t_i\rightarrow t_{i+1}}}_{p,q}$ as the growth rate $g_p$ of each cell $p$, and propagate cell $p$ from time $t_i$ to $t_{i+1}$ using the multiplier $g_p^{t_{i+1}-t_i} / \sum_p g_p^{t_{i+1}-t_i}$. 
This modifies the gene expression of each cell, and WOT restarts the whole algorithm iteratively with the computed growth rate.
WOT's strategy discriminates WOT from OT solvers and integrates time and biological information into its algorithm.

We view the optimal transport plan as a transition matrix between cells in $X^{t_i}$ and $X^{t_{i+1}}$, and reconstruct the cell evolutionary trajectory $V_p$ by iteratively finding the most probable cell at the next time step from the transition matrix.
\subsection{Neural Relational Inference (NRI)}
A graph neural network (GNN) is a type of neural network that can operate on graph-structured data and maintain the node invariant property in its message-passing mechanism.
Denote $h_r^l$ as the node embedding of node $v_r\in\mathcal{V}$ at layer $l$ and $h_{r,s}^l$ as the edge embedding of edge $e_{r,s}\in\mathcal{E}$ at layer $l$.
The message-passing mechanism in GNN is defined as:

\begin{align}
    \text{Node embedding:} \quad h_r^1 &= f_{emb}(v_r) \\
    \text{Node-to-edge:} \quad h_{r,s}^l &= f_e^l(h_r^l,h_s^l) \\
    \text{Edge-to-node:} \quad h_r^{l+1} &= f_v^l\left(\sum\nolimits_{s\in\mathcal{N}(r)} h_{r,s}^l\right),
\end{align}

where $f_{emb}$ is the node embedding network, $f_e^l$ is the node-to-edge message-passing network at layer $l$, $f_v^l$ is the edge-to-node message-passing network at layer $l$, and $\mathcal{N}(p)$ is the set of neighbors of node $v_p$.

NRI is a structural inference model that uses GNNs to infer the underlying structure of a dynamical system from observed trajectories.
It has a VAE architecture.
Encoder in NRI uses $f_{emb}\rightarrow f_e^1\rightarrow f_v^1\rightarrow f_e^2$ to send and encode the node feature to its edge embeddings $Z\in\mathbb{R}^{g\times g}$, where $Z_{r,s}$ is the Gumbel softmax of $h^2_{r,s}$.
The network functions $f_{(\dots)}$ are multilayer perceptron (MLP) or CNN.
The decoder of NRI predicts the expression level of gene $s$ in the trajectory $V_p$ at the next time step $i+1$ through:

\begin{align}
V_{p,r}^{i+1} = V_{p,r}^{i} + \tilde{f}_v\left(\sum\nolimits_{s\neq r}Z_{r,s}\tilde{f}_e(V_{p,r}^{i},V_{p,s}^{i})\right),
\end{align}

where $\tilde{f}_v$ is the edge-to-node message-passing network and $\tilde{f}_e$ is the node-to-edge message-passing network.
The network function $\tilde{f}_v$ can be a recurrent neural network (RNN) or an MLP.
RNN decoder is designed for dynamic systems which is not Markovian.
Upon the model converges, the softmax of encoder output $h^2_{r,s}$ gives the probabilities of edges existence of $e_{r,s}$

Wang and Pang~\cite{wang2022isidg} notice that a general GNN composites the operations $f_{emb}\rightarrow f_e^1\rightarrow f_v^1$, and thus the NRI encoder can be written as 
\begin{align}
h_{r,s}^2 = f_e^2(f_{\text{GNN}}(v_r,v_s))
\end{align},
where $f_{\text{GNN}}$ can be any state-of-the-art GNN architecture.
They~\cite{wang2022isidg} show that graph isomorphism network (GIN)~\cite{xu2019gin} is the best NRI encoder choice on a simulated scRNA-seq dataset, and adding hybrid loss can improve model performance and control the smoothness of inferred graph.
\section{Our Method}
\label{sec:method}
\begin{figure}[!t]
    \includegraphics[width=\textwidth]{"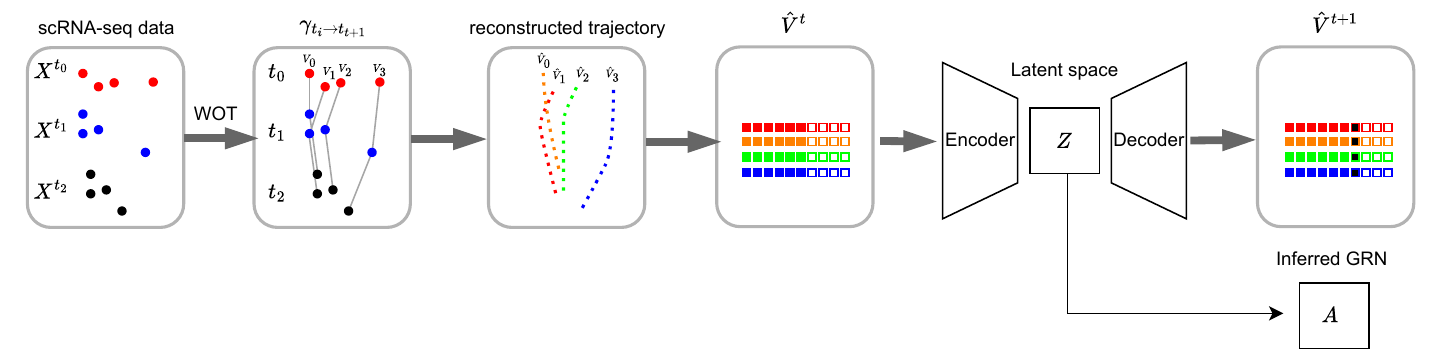"}
    \caption{Overview of the proposed pipeline. With gene expression matrixes at multiple time points as input, we first connect cells at consecutive time points using WOT. The reconstructed cell evolutionary trajectories are then fed into the NRI model. The inferred GRN is the output of the trained NRI encoder.}
    \label{fig:method}
\end{figure}

We reconstruct the trajectories using WOT and feed them to the NRI model to infer the underlying GRN.
The overall method is illustrated in Fig.~\ref{fig:method}.

We view the optimal transport plan as a transition matrix between cells in $X^{t_i}$ and $X^{t_{i+1}}$.
For each cell $p$ at time $t_0$, we reconstruct the cell evolutionary trajectory $V_p=\{V_p^{t_i}\}_{i=0}^{k-1}$ by iteratively finding the next most probable cell at the next time step through the transition matrix $\gamma_{t_i\rightarrow t_{i+1}}$. 
This approach benefits from the fast computation of transition matrixes and the natural integration of temporal information and biological considerations from WOT.
We then train the NRI model with the reconstructed trajectories.
\section{Experiments}
\label{sec:exp}
To evaluate the performance of our proposed method, we conduct experiments on synthetic datasets, namely mCAD and VSC.
\subsection{Experiment Setup}
\label{ssec:preprocessing}
We extracted two curated GRNs from the popular BEELINE~\cite{pratapa2020benchmarking} benchmark, namely mCAD and VSC, each including 5 and 8 genes, respectively.

For each curated GRN, we used BoolODE~\cite{pratapa2020benchmarking}, a stochastic differential equation-based scRNA-seq data simulator, to simulate 8,000 cells evolving in 1,000 time steps governed by its corresponding GRN.
The evolution of cells creates a three-dimensional array, storing 8,000 trajectories of length 1,000 with dimensions equal to their underlying GRN.
We only retained 7 time steps, $t=0,50,100,400,600,650,1000$, in the trajectories to minimic the sparse and irregularly sampled experimental gene expression data.
We split the trajectories into a train set for hyperparameter tuning and a test set for evaluating, with a ratio of $6:2$.
We shuffled the cell at each time step to break the trajectories.

We then apply the trajectory reconstruction method mentioned in Section~\ref{sec:method} to the datasets.
Each trajectory at each time point is a multidimensional vector representing the gene expression levels of a cell.
The accuracy of trajectory reconstruction is explored in Section~\ref{ssec:evaluation_traj_reconstruction}.
We tune the hyperparameters in WOT and NRI on the train set and report the model performance on the test set.

Our NRI model is trained with a GIN encoder with a 0.3 dropout rate, an MLP decoder with a 0.5 dropout rate, a smoothness penalty of -500, and trained with 500 epochs.
The model is trained with ten random seeds on an NVIDIA A100 40GB HBM GPU.
Each evaluation metric's mean and standard deviation are reported.
\subsection{Evaluation on Trajectory Reconstruction}
\label{ssec:evaluation_traj_reconstruction}
\begin{figure}[!t]
    \includegraphics[width=\textwidth]{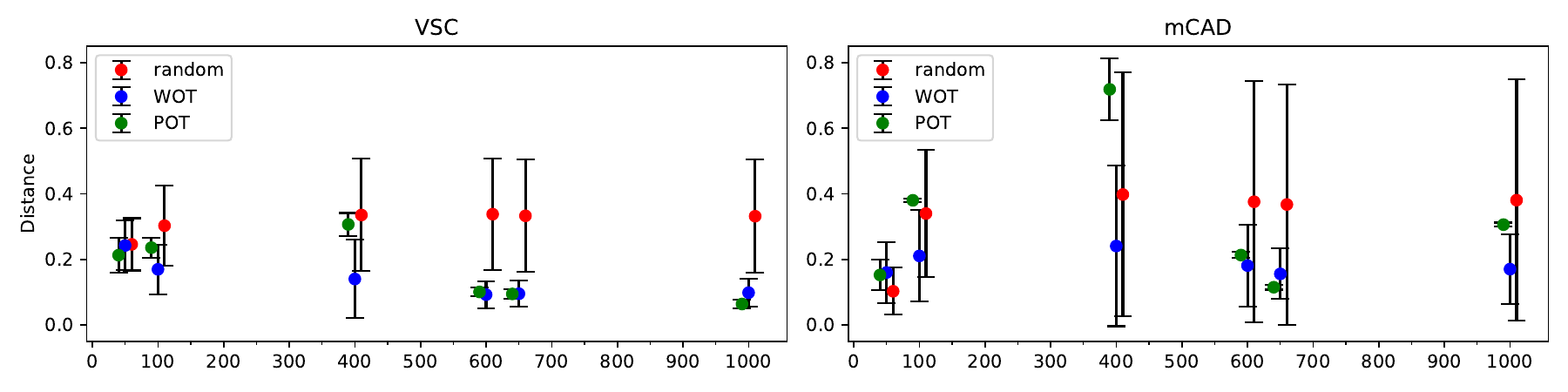}
    \caption{Comparison of normalized trajectory reconstruction error from time $i$ to $i+1$ in different GRNs between WOT, OT and random.}
    \label{fig:reconstruction_error}
\end{figure}

We first validate the performance of trajectory reconstruction using WOT and naive OT, as accurate reconstructed trajectories are fundamental in our study.
After reconstructing trajectories, we measure the normalized reconstruction error by $\frac{1}{g}\|\hat{V}^t_p-V^t_p\|_2$, where $\hat{V}^t_p,V^t_p$ are the predicted and actual gene expression vectors of a particular cell $p$ at time $t$, $g$ is the number of genes.
The normalized reconstruction error of different GRNs using WOT, naive OT and no reconstruction (random) are shown in Fig.~\ref{fig:reconstruction_error}.

We observed that both OT and WOT can significantly reduce the reconstruction error to around 0.2 in gene expression value.
While OT is more stable in reconstruction performance, WOT can achieve a lower reconstruction error.
This experiment shows how integrating temporal information and biological considerations in the WOT algorithm helps reconstruct cellular trajectories.
\subsection{Evaluation Metrics on Structural Inference}
\label{ssec:evaluation_metrics}
Our pipeline outputs an adjacency matrix $A\in\mathbb{R}^{g\times g}$, where $A_{r,s}$ is the probability of the existence of directed interaction from gene $r$ to gene $s$.

We evaluate the performance of the inferred GRN using the following metrics: area under the receiver operating characteristic curve (AUROC), area under the precision-recall curve (AUPRC), and the early precision rate (EPR), defined as the ratio of true positive edges against a random predictor in their top-$k$ predicted edges, where $k$ is the total number of edges in the ground-truth network.

In this paper, we primarily use AUROC to guide our hyperparameter tuning, maintaining consistency with previous research in the GRN inference field while reporting AUPRC and EPR values as references.

\subsection{Results}
\label{ssec:results}
We compare our model with the following baselines:
\begin{itemize}
    \item ppcor~\cite{kim2015ppcor}, which computes the partial correlation matrix as the GRN;
    \item ARACNe~\cite{margolin2006aracne}, which treats the mutual information (MI) as the strength of gene interaction and removes extra edges using data processing inequality;
    \item CLR~\cite{faith2007clr}, which also treats the MI as gene interaction strength but identifies MI outliers as true interaction;
    \item PIDC~\cite{chan2017pidc}, which employ partial information decomposition~\cite{williams2010nonnegative} to decompose MI and only retain the unique information as the gene interaction;
    \item GENIE3~\cite{huynh2010genie3}, a random forest model trained by predicting one gene by other genes and unboxing the gene importances as the gene interaction strength;
    \item GRNBOOST2~\cite{moerman2019grnboost2}, a model similar to GENIE3 but replacing the random forest model into a gradient boosting machine;
    \item dynGENIE3~\cite{huynh2018dyngenie3}, a dynamic version of GENIE3 that models the cell dynamic by ODE and the random forest model;
    \item XGBGRN~\cite{ma2020inference}, a dynamic version of GENIE3 that models the cell dynamic by ODE and the XGBoost model.
\end{itemize}

ppcor is a classical statistical method which only relies on partial correlation.
ARACNe, CLR and PIDC are MI-based methods which distinguish real gene interactions from noise.
GENIE3, GRNBOOST2, dynGENIE3 and XGBGRN are machine learning-based methods, while the latter two can be further labelled as dynamic modelling as they both utilize temporal information.
All baselines can only give a numeric value for each pair of genes indicating the gene relation, but manual thresholding is required to conclude the existence of regulation.

\begin{table}[!t]
\centering
\caption{Average AUROC (in \%) and AUPRC (in \%) of selected GRN inference models on two synthetic datasets in ten runs.
         The starred (*) model indicates that the model is deterministic and that no standard deviation is provided.
         The best performance with respect to each metric is highlighted by boldface.}
\label{tab:result_no_epr}
\begin{tabular}{c|ll|ll|}
\multicolumn{1}{l|}{} & \multicolumn{2}{c|}{AUROC (in \%)}                               & \multicolumn{2}{c|}{AUPRC (in \%)}                            \\ \hline
\multicolumn{1}{}{} & \multicolumn{1}{|c|}{mCAD}       & \multicolumn{1}{c|}{VSC}        & \multicolumn{1}{c|}{mCAD}       & \multicolumn{1}{c|}{VSC}      \\ \hline \hline
ppcor*                & 54.87                          & 68.37                          & 59.04                          & 53.84                       \\
ARACNe*               & 52.27                          & 75.99                          & 59.98                          & 54.19                       \\
CLR*                  & 61.69                          & 74.90                          & 61.18                          & 52.76                       \\
PIDC*                 & 55.19                          & 81.16                          & 56.38                          & 67.21                       \\
GENIE3               & 55.58$\pm$0.43          & \textbf{83.58$\pm$0.49} & 64.30$\pm$0.41          & \textbf{74.88$\pm$0.38} \\
GRNBOOST2            & 52.08$\pm$2.56          & 81.28$\pm$2.20          & 58.23$\pm$2.86          & 64.48$\pm$4.97          \\
dynGENIE3            & 24.16$\pm$0.76          & 71.55$\pm$1.04          & 45.63$\pm$1.45          & 34.67$\pm$0.51          \\
XGBGRN               & 44.81$\pm$0.00          & 78.50$\pm$0.00          & 50.81$\pm$0.00          & 63.47$\pm$0.00          \\
WOT+NRI (Ours)       & \textbf{73.96$\pm$3.08} & 81.39$\pm$7.90          & \textbf{79.31$\pm$4.68} & 66.58$\pm$6.18          
\end{tabular}
\end{table}

\begin{table}[!t]
    \centering
    \caption{Average EPR value of selected GRN inference models on two synthetic datasets in ten runs.
             The starred (*) model indicates that the model is deterministic and that no standard deviation is provided.
             The best performance with respect to each metric is highlighted by boldface.}
    \label{tab:result_epr}
    \begin{tabular}{c|ll|}
    \multicolumn{1}{l|}{} &  \multicolumn{2}{c|}{EPR}                                         \\ \hline
    \multicolumn{1}{}{} &    \multicolumn{1}{|c|}{mCAD}       & \multicolumn{1}{c|}{VSC}        \\ \hline \hline
    ppcor*                &  0.893                          & 1.707                          \\
    ARACNe*               &  1.020                          & 2.276                          \\
    CLR*                  &  1.276                          & 2.276                          \\
    PIDC*                 &  1.276                          & 2.276                          \\
    GENIE3               &   1.020$\pm$0.00          & \textbf{2.844$\pm$0.00} \\
    GRNBOOST2            &   1.020$\pm$0.00          & 2.190$\pm$0.22          \\
    dynGENIE3            &   0.753$\pm$0.04          & 1.422$\pm$0.00          \\
    XGBGRN               &   0.893$\pm$0.00          & 1.991$\pm$0.00          \\
    WOT+NRI (Ours)       &   \textbf{1.339$\pm$0.09} & 2.588$\pm$0.37         
    \end{tabular}
    \end{table}

A performance comparison of our model with the baselines is shown in Table~\ref{tab:result_no_epr} and ~\ref{tab:result_epr}.
Our method and GENIE3 are the best performers in all metrics for the mCAD and VSC datasets, respectively.
All methods except our approach and CLR perform poorly in the mCAD dataset with AUROC less than 56\%. 
Most of the methods have high AUROC in the VSC dataset, with half of the models having an AUROC value exceeding 80\%. 
Notably, dynamic models dynGENIE3 and XGBGRN have a huge performance gap between the two datasets.
\section{Discussion}
\label{sec:discussion}
In this work, we have demonstrated the capability of our method to infer two GRNs from simulated single-cell gene expression dataset accurately.
Our approach has two advantages.

First, our approach can infer asymmetric GRN from temporal datasets without thresholding to determine the existence of interactions.
As mentioned in Section~\ref{ssec:results}, only two baseline methods, dynGENIE3 and XGBGRN, utilize temporal information during structural inference, and both of them exhibit huge discrepancies in model performance between mCAD and VSC datasets.
Besides, all studied baselines require thresholding to conclude the existence of gene interactions.
Thus, our approach filled the research gap in GRN inference by utilizing temporal information with no thresholding because our approach can directly give the probabilities of interactions between genes.

Second, our approach paves the way to harvest the powerful trajectory-based structural inference models such as iSIDG~\cite{wang2022isidg}, RCSI~\cite{wang2023rcsi} and GDP~\cite{pan2023graph}.
Our approach reconstructs the destroyed trajectory information, and we can replace the NRI with other trajectory-based structural inference models.
iSIDG and RCSI have shown promising results on GRN inference in synthetic datasets with trajectory information.
Our work can bridge two scientific communities together and import new structural inference models into the GRN inference society.

Despite the attractive performance, our approach faces two limitations.

First, we have only tested our approach's performance on a synthetic dataset, and the performance on a real dataset is unexplored.
Although BoolODE uses chemical Langevin equations in cell dynamic modelling, some noise sources in the real world are ignored, such as doublets, missing reads during assaying, and mRNA degradation before the polymerase chain reaction amplification.
Also, we cannot observe the zero inflation in the synthetic dataset, which is common in real-world datasets.

Besides, we only chose two small-scale curated GRNs for the experiment due to time and computation cost constraints.
Experimenting with more curated GRNs facilitates objective and unbiased evaluation of our pipeline.
Also, we should aim for GRN with a larger scale.
The real-world dataset covers thousands of protein-encoding genes, and NRI and its variants cannot tackle a network of this scale because of limited GPU memory.
Our work serves as a proof of concept, and we leave an unbiased evaluation and scalability analysis for our future work.
\section{Conclusion and Future Work}
\label{sec:conclusion}
We have introduced a novel pipeline for inferring the GRN from single-cell gene expression data by integrating the optimal transport algorithm with the structural inference model.
We applied the WOT algorithm to find the optimal mapping between the cells at different time points and used the NRI model to infer the GRN from the reconstructed trajectories.
We have shown that our method outperforms most of the state-of-the-art GRN inference methods on both curated datasets.
Our approach can also adopt future trajectory-based structural inference methods to replace NRIs.
In future, we plan to perform more case studies of inferring GRNs from both synthetic and real-world single-cell data to extensively evaluate our method and compare it to other GRN inference methods, especially those based on deep learning.

However, through the case study on two curated datasets, we also identified several limitations of our approach, including the size and number of chosen curated GRNs and the lack of experiments on real-world datasets.
We leave unbiased training and evaluation for our future work.
Researchers can extend our work in the following aspects.

From the NRI configuration perspective, we can train the NRI model with three edge types (activation, inhibition, and no interaction), which allows us to distinguish between activation and inhibition interactions.
Besides, we can attach additional features to the reconstructed trajectories, such as RNA velocity and the single-cell assay for transposase-accessible chromatin using sequencing (scATAC-seq) data, forcing NRI to incorporate this biological information.

From the model perspective, we can leverage popular Neural ODE models~\cite{tong2020trajectorynet,huguet2022mioflow,sha2024tigon} to reconstruct cellular trajectories.
Neural ODE directly models the cell dynamics and outputs a velocity field that transports cell distributions between consecutive time points.
Integrating such a velocity field could reconstruct continuous cell trajectories at any sampled cell.
Although current models suffer from scalability~\cite{sha2024tigon} and convergence~\cite{tong2020trajectorynet} issues, it is a promising field potentially discovering gene interactions at higher orders.
We expect that our work will inspire future research in GRN inference and contribute to the development of more accurate GRN inference methods.

The code and datasets used in this study are publicly available at \url{https://github.com/1250326/Integrating-OT-and-Structural-Inference-Models-for-GRN-Inference-from-Single-cell-Data}.

\begin{credits}
\subsubsection{\ackname} Authors Tsz Pan Tong and Jun Pang acknowledge financial support of the Institute for Advanced Studies of the University of Luxembourg through an Audacity Grant (AUDACITY-2021).
The training and evaluation were performed on the Luxembourg national supercomputer MeluXina.
The authors gratefully acknowledge the LuxProvide teams for their technical support.

\subsubsection{\discintname}
The authors have no competing interests to declare that are relevant to the content of this article.
\end{credits}

%
%
%
%

\bibliography{bib}
\bibliographystyle{splncs04}





\end{document}